\documentstyle[sprocl,epsfig]{article}

\def\Journal#1#2#3#4{{#1} {\bf #2}, #3 (#4)}

\def\NPA{{\em Nucl. Phys.} A}
\def\NPB{{\em Nucl. Phys.} B}
\def\PLB{{\em Phys. Lett.}  B}
\def\PRL{\em Phys. Rev. Lett.}

\def\PRC{{\em Phys. Rev.} C}

\def\PR{{\em Phys. Rep.}}
\def\JPG{{\em J. Phys.} G}
\def\EPJC{{\em Eur. Phys. J.} C}

\def\beq{\begin{equation}}
\def\eeq{\end{equation}}
\def\bea{\begin{eqnarray}}
\def\eea{\end{eqnarray}}
\def\eg{{\it e.g.}}
\def\ie{{\it i.e.}}
\def\etal{{\it et al.}}
\newcommand{\tave}[1]{\langle\!\langle{#1}\rangle\!\rangle}
\newcommand{\bce}{\begin{center}}
\newcommand{\ece}{\end{center}}

\begin{document}

\title{Duality and Chiral Restoration from Dilepton Production 
in Relativistic Heavy-Ion Collisions}

\author{Ralf Rapp}

\address{Department of Physics and Astronomy, State University of New York,\\
Stony Brook, NY 11794-3800, U.S.A. \\email: rapp@tonic.physics.sunysb.edu}

\maketitle\abstracts{We discuss the recent status in the theoretical
understanding of dilepton production in central heavy-ion reactions
with the Pb-beam at the full CERN-SpS energy of 158~AGeV. In the low-mass
region ($M\le$~1~GeV) a strong broadening of the vector meson resonances 
in hot and dense matter (especially for the $\rho$ meson) entails 
thermal dilepton rates very reminiscent to perturbative $q\bar q$ 
annihilation close to the expected phase
boundary of the chiral symmetry restoring transition. A
consistent description of the experimentally observed enhancement
at both low and intermediate masses (1.5~GeV~$\le M \le$~3~GeV) in terms
of thermal radiation from an expanding fireball can be obtained.}

\section{Introduction}
The basic objective of the relativistic heavy-ion program is the 
investigation of the many-body properties of Quantum Chromodynamics
associated with, \eg, its phase diagram.
Two of the main nonperturbative features characterizing the ground 
state (vacuum) of the theory -- confinement and the spontaneous breaking
of chiral symmetry -- are expected to cease in possibly common phase 
transitions at finite temperature and/or density (see
Ref.~\cite{Maw99} for recent results from lattice simulations).  
Due to their negligible rescattering dileptons ($l^+l^-=e^+e^-,
\mu^+\mu^-$) can in principle directly probe the highest excitation 
zones formed in the early stages of central nucleus-nucleus collisions. 
However, the measured spectra also contain radiation from later stages
as well as the decay contributions from hadrons in the final state
(the so-called 'cocktail'). Whereas
the latter can be inferred from measured hadron spectra, the 
radiation from the fireball requires both the modeling of the
space-time evolution of a heavy-ion reaction and the evaluation of 
dilepton production rates in strongly interacting matter.

In the low-mass region (LMR, $M\le$~1~GeV), 
where hadronic properties are governed by 
the spontaneous breakdown of chiral symmetry, one hopes to witness its
restoration through medium effects visible in the direct decays of the light
vector mesons, $V\to l^+l^-$ ($V=\rho, \omega, \phi$). Definite conclusions, 
however, require the simultaneous treatment of their chiral partners:  
vector and  axialvector mesons have to become degenerate
at the transition. The central question then is, {\em how} this is realized,
\eg, do their masses merge to zero? Do their widths diverge? etc.
(see Ref.~\cite{RW00} for a recent review).    

The intermediate-mass region (IMR, 1.5~GeV~$\le M \le$~3~GeV)
has long been proposed as a suitable window to observe thermal radiation
from an equilibrated Quark-Gluon Plasma (QGP)~\cite{Sh80}: 
at the larger masses the thermal signal is more sensitive
to higher temperatures, whereas light-hadron decay contributions are 
concentrated at low masses 
and initial hard processes such as Drell-Yan annihilation, which dominate
in the high-mass region ($M\ge 4$~GeV), might be sufficiently suppressed 
towards the IMR. 
  
Here, we will try to give a unified description of both low- and 
intermediate-mass dilepton measurements at the CERN-SpS, in connection   
with  possible indications of phase transition signatures.
We first discuss various approaches to assess 
equilibrium dilepton production rates and their medium effects
(Sect.~2), and then proceed to the
application for evaluating dilepton observables coupled with a simplified  
space-time description of heavy-ion reactions (Sect.~3).

\section{Electromagnetic Current Correlator and Medium Effects}
\begin{figure}[!htb]
\bce
\epsfig{file=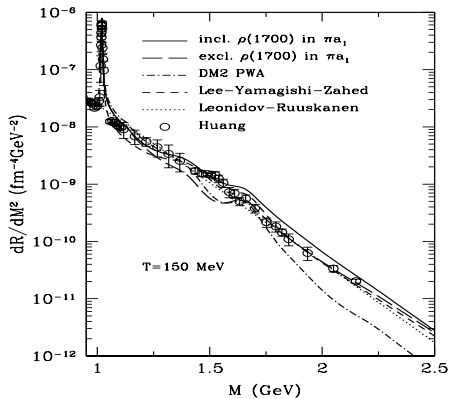,width=8cm,height=6cm}
\ece
\vspace{-0.9cm}
\caption{A compilation of 3-momentum integrated dilepton production rates 
in the IMR~\protect\cite{LG98}. The open points are 
inferred~\protect\cite{Huang} via empirical information from 
$\sigma(e^+e^-\to hadrons)$ including the mixing 
effect, Eq.~(\protect\ref{vamix}), the dotted line corresponds to perturbative
$q\bar q$ annihilation~\protect\cite{LR98} and all other curves are variants 
of multi-hadronic calculations~\protect\cite{LG98,LYZ98}.}
\label{fig_imr}
\end{figure}

The thermal dilepton production rate from a hot and
dense medium can be decomposed as
\beq
\frac{dN_{l^+l^-}}{d^4xd^4q}=L_{\mu\nu}(q) \ 
W^{\mu\nu}(M,\vec q ;\mu_B, T) \ , 
\eeq
where $L_{\mu\nu}(q)$ is the standard lepton tensor, and
the hadron tensor  $W^{\mu\nu}$ contains all the non-trivial information
on the hadronic medium of temperature $T$ and baryon chemical potential
$\mu_B$. It is defined via
the thermal expectation value of the electromagnetic (e.m.) current-current
correlator, 
\bea
W^{\mu\nu}(q) &=& -i \int d^4x \ e^{-iq\cdot x} \
\tave{j^\mu_{\rm em}(x) j^\nu_{\rm em}(0)}_T
\nonumber\\
 &=& \frac{-2}{\exp(q_0/T)-1} \ {\rm Im} \Pi^{\mu\nu}_{\rm em}(q) \ .
\eea
Depending on the invariant masses probed, the e.m.~correlator can be
described by either using hadronic degrees of freedom (saturated
by vector mesons within the well-established vector dominance model (VDM))
or the (perturbative) quark-antiquark vector correlator, \ie,
\beq
{\rm Im} \Pi_{\rm em}^{\mu\nu}  = \left\{
\begin{array}{ll}
 \sum\limits_{V=\rho,\omega,\phi} ((m_V^{(0)})^2/g_V)^2 \
{\rm Im} D_V^{\mu\nu}(M,\vec q;\mu_B,T) & , \ M \le M_{dual}
\vspace{0.3cm}
\\
(-g^{\mu\nu}+q^\mu q^\nu/M^2) \ (M^2/12\pi) \ N_c
\sum\limits_{q=u,d,s} (e_q)^2  & , \ M \ge M_{dual} \ 
\end{array}  \right.
\label{ImPiem}
\eeq
(Im$D_V$: vector meson spectral function).
In vacuum the transition region is located at a 'duality threshold'
of $M_{dual}\simeq 1.5$~GeV, as marked by the inverse process
of $e^+e^-$ annihilation into hadrons. Indeed, above $M_{dual}$ 
the results from rather 
complete hadronic calculations closely coincide with the simple 
perturbative expression, 
\beq
\frac{d^8N_{l^+l^-}}{d^4xd^4q} = \frac{\alpha^2}{4\pi^4} f^B(q_0;T)
\sum\limits_{q=u,d,s} (e_q)^2 \  , 
\label{pert}
\eeq
cf.~Fig.~\ref{fig_imr}. We note that in the IMR  both temperature and 
density corrections, being of order $O(T/M)$ and $O(\mu_q/M)$,  
are negligible under the conditions probed at the SpS.

The situation is more involved in the LMR. Over the last years
large efforts have been undertaken to assess medium effects in the 
vector correlator. Model independent results, based on current algebra
together with a low-temperature (-density) expansion, have shown
that (in the chiral limit) the leading effect is a mere mixing between 
the vector and axialvector degrees of freedom~\cite{DEI90,Krippa}, 
\bea
\Pi^{\mu\nu}_V(q) &=& (1-\varepsilon) \ \Pi^{\circ\mu\nu}_V(q)
+\varepsilon \ \Pi^{\circ\mu\nu}_A(q)
\nonumber\\
\Pi^{\mu\nu}_A(q) &=& (1-\varepsilon) \ \Pi^{\circ\mu\nu}_A(q)
+\varepsilon \ \Pi^{\circ\mu\nu}_V(q) \
\label{vamix}
\eea  
with the mixing parameter $\varepsilon=T^2/6f_\pi^2$.
Extrapolating to chiral restoration ($\varepsilon=1/2$) one finds 
$T_c^\chi\simeq 160$~MeV (or $\rho_c^\chi\simeq2.5~\rho_0$~\cite{Krippa}), 
which is not 
unreasonable. More surprisingly, the fully mixed vector correlator leads to
a dilepton production rate which agrees well 
with perturbative QGP radiation 
down to the $\phi$ resonance~\cite{Huang}, cf.~open points and 
dotted line in Fig.~\ref{fig_imr} (note that the mixing does not affect
the resonance structures in $\Pi_V^\circ$). 
Thus one concludes that in the 1~GeV~$<M<$~1.5~GeV region
chiral restoration is realized through a lowering of the in-medium
duality threshold, being a 'weak' temperature effect. 

The investigation of in-medium modifications of the low-lying   
vector resonances has been pursued in a variety of frameworks~\cite{RW00}. 
Within the mean-field approximation, and using arguments of scale
invariance of the QCD Lagrangian at finite temperature/density, 
Brown and Rho predicted the $\rho$ and $\omega$ meson masses 
to follow a universal 'scaling law'~\cite{BR91}, \ie, 
$m_{\rho,\omega}(\rho,T)$ drop (in line with other non-Goldstone hadrons) 
in accord with the 
vanishing the pion decay constant $f_\pi$, which is one of the
order parameters of chiral restoration. 

\begin{figure}[!htb]
\epsfig{file=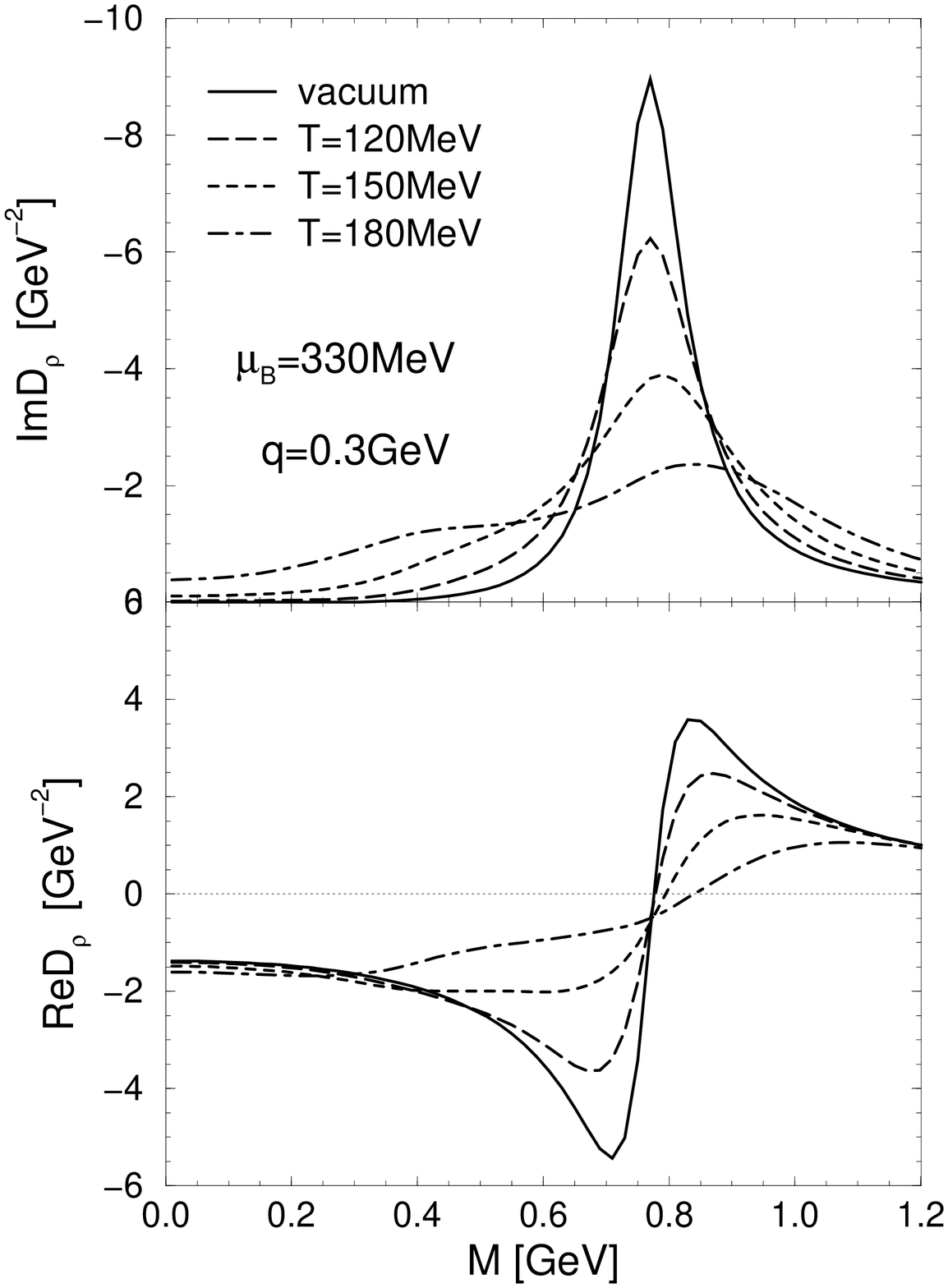,width=5.5cm}
\epsfig{file=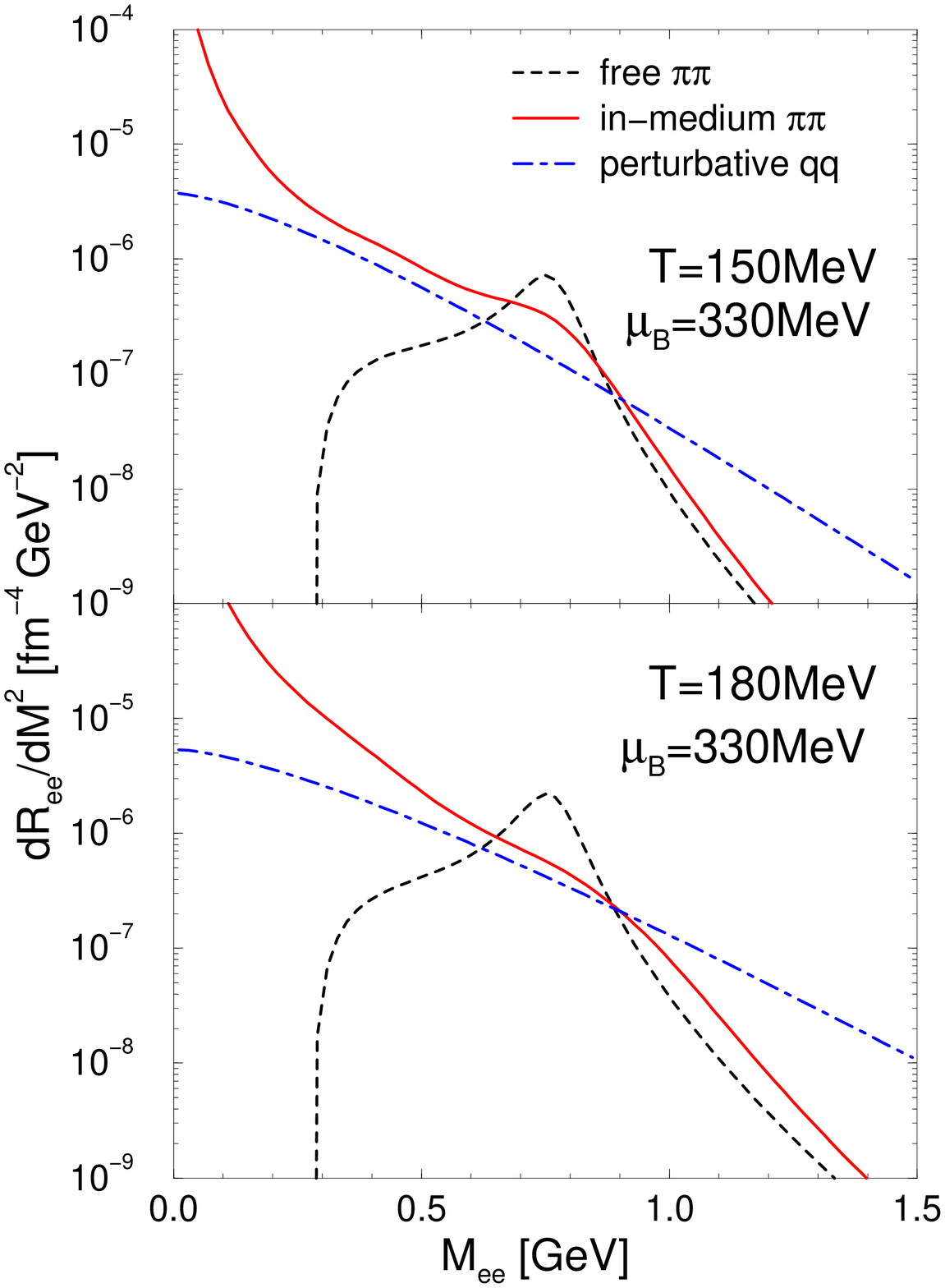,width=5.5cm}
\caption{In-medium $\rho$ propagator (left panel) and pertinent dilepton 
production rates (right panel, together with pQCD results) 
as emerging from a hadronic many-body 
calculation~\protect\cite{RW99}.}
\label{fig_lmr}
\end{figure}
Other approaches aim at a proper description of the in-medium vector spectral
functions based on  phenomenologically well-established interactions
with surrounding matter particles such as pions or nucleons in connection with
standard many-body techniques. These calculations
are based on chirally and gauge invariant couplings where the
associated parameters are  constrained by partial (hadronic and
electromagnetic) decay widths of the excited resonances
(\eg, $a_1\to \pi\rho, \pi\gamma$) or, if available, more comprehensive
information as encoded, \eg, in photoabsorption spectra on 
nucleons~\cite{KW96} and nuclei~\cite{RUBW}.
An example of such a calculation~\cite{RW99} is displayed in 
Fig.~\ref{fig_lmr}.
As a result of multiple interactions in hot/dense matter 
the $\rho$ spectral function (upper left panel) undergoes a strong
broadening thereby losing its quasiparticle nature (as is also
evident from the flattening of the real part, cf.~lower left panel). 
At the highest temperatures/densities the corresponding dilepton 
production rates (right panel) exhibit a remarkable reminiscence
to perturbative $q\bar q$ annihilation in the 0.5--1~GeV region
(above, the lack of 4-pion states in the $\rho$ propagator leads
to the falloff of the full lines; cf.~Fig.~\ref{fig_imr} for the 
more complete description in the IMR). This may be interpreted 
as a further penetration of the in-medium duality threshold
down to rather low masses of 0.5~GeV, this time caused by strong
rescattering effects resummed in the Dyson equation for the $\rho$ 
propagator. 

\section{Dilepton Spectra in Pb(158~AGeV)-Induced Reactions}
To compare with experiment, the dilepton rates discussed in the
previous section have to be folded over a realistic space-time
evolution~\cite{CB99} of a given heavy-ion reaction. We here
employ a thermal fireball expansion based on a resonance gas equation 
of state. The latter is constructed in accord with recent analysis
on the {\it chemical} freezeout at SpS energies~\cite{chem}, 
\ie, the stage
where particle abundances are frozen.  Entropy as well as baryon-number 
conservation determine the thermodynamic trajectory towards {\it thermal}
freezeout, where rescattering ceases. In addition, effective 
pion-number conservation induces the build-up of finite pion 
chemical potentials~\cite{Bebie}, reaching $\mu_\pi=60-80$~MeV. 
The thermal radiation contribution
from the fireball then takes the form
\beq
\frac{dN_{l^+l^-}^{th}}{dM}=\frac{\alpha^2}{\pi^3 M}
\int\limits_0^{t_{fo}} dt \ V_{FC}(t) \int \frac{d^3q}{q_0} 
\ f^B(q_0;T,\mu_\pi) \ 
{\rm Im} \Pi_{\rm em} \ {\rm Acc}(M,q_t,y) \ ,
\eeq
where ${\rm Acc}(M,q_t,y)$ accounts for the specific detector acceptance and
$V_{FC}(t)$ describes the (cylindrical) volume expansion.  The thermal Bose 
factor $f^B$ incorporates $\mu_\pi>0$ in the 
hadronic phase, which in Boltzmann approximation amounts to an
enhancement  factor $(e^{\mu_\pi/T})^n$. In the LMR, where two-pion
annihilation prevails, $n=2$, whereas in the IMR $n=4$ corresponding
to the dominant $\pi a_1$ channel~\cite{LG98}.   
\begin{figure}[!htb]
\vspace{-0.8cm}
\begin{minipage}[t]{5.8cm}
\bce
\vspace{1.1cm}
\epsfig{file=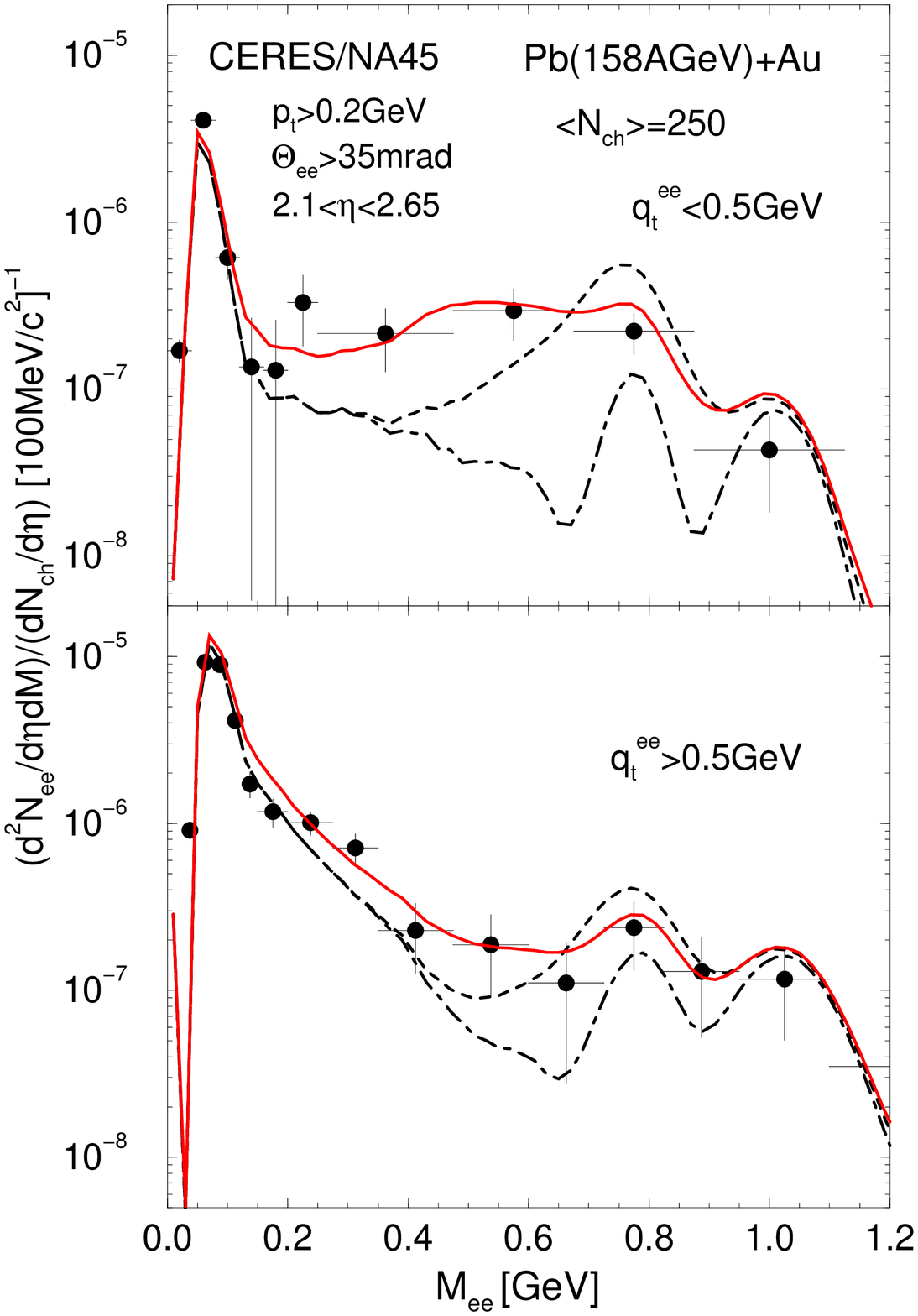,width=6.2cm}
\ece
\end{minipage}
\begin{minipage}[t]{5.2cm}
\epsfig{file=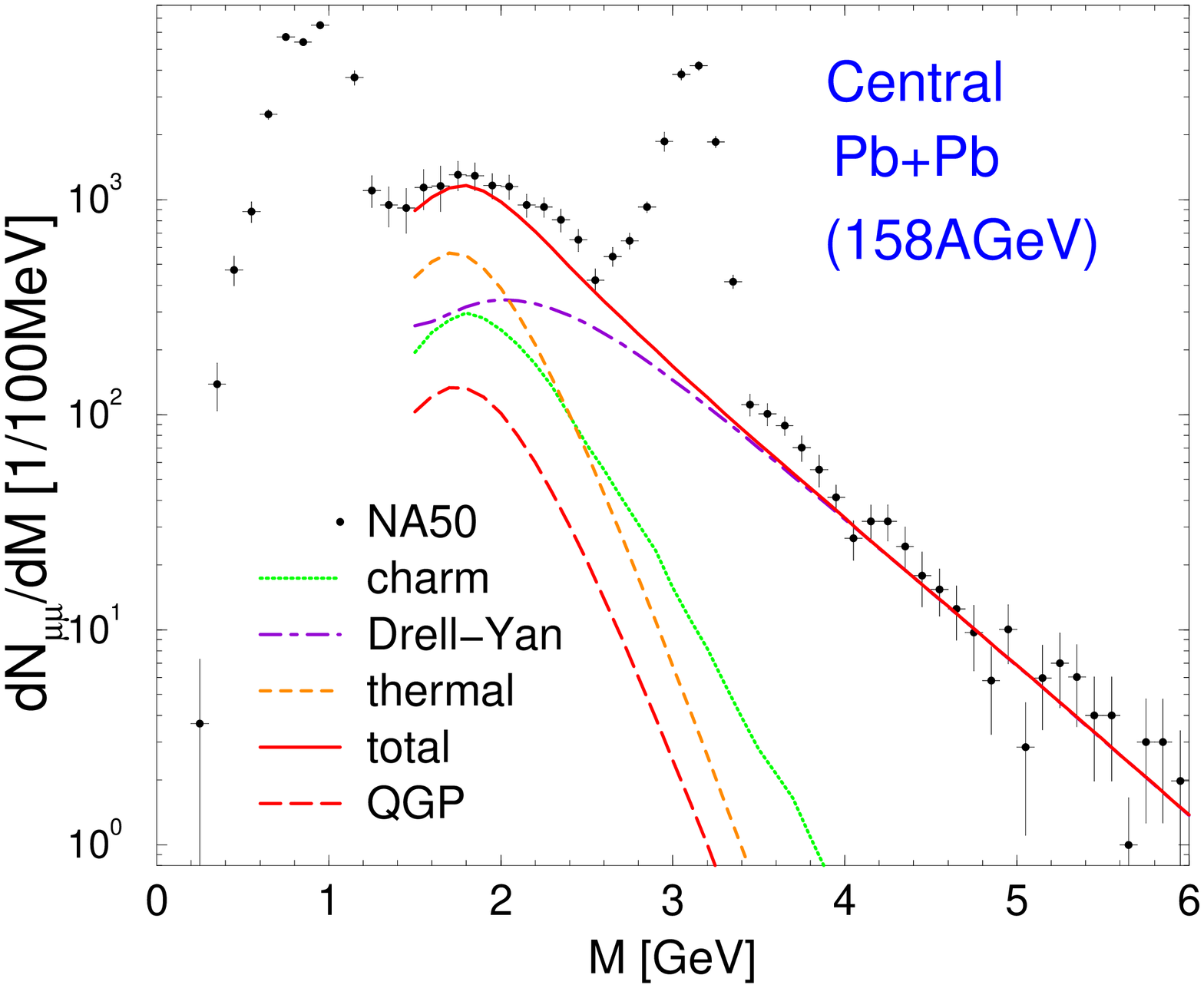,width=4.9cm,angle=-90}
\epsfig{file=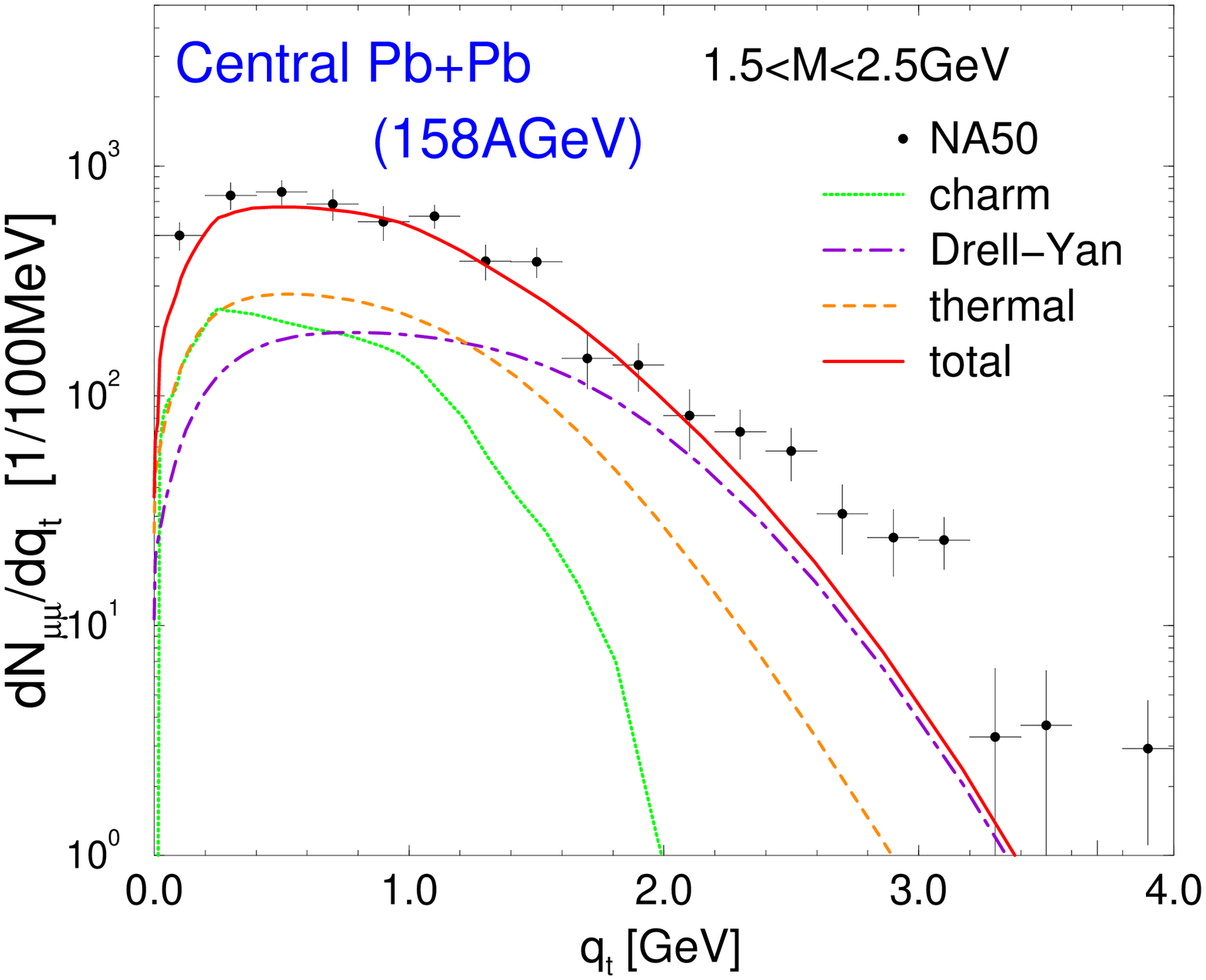,width=4.9cm,angle=-90}
\end{minipage}
\caption{Dilepton Spectra at the CERN-SpS. Left panel: LMR, split into
two $q_t$-bins, as measured by CERES/NA45~\protect\cite{ceres}
 in 30\% central Pb+Au 
(dashed-dotted lines: hadronic cocktail, 
dashed line: cocktail + free $\pi\pi$ annihilation, solid lines:
cocktail + in-medium $\pi\pi$ annihilation). Right panel: IMR as
measured by NA50~\protect\cite{na50} in central Pb+Pb (upper/lower panel: 
M-/$q_t$-spectra; the long-dashed line in the upper panel 
constitutes the contribution from the QGP phase assuming
a critical temperature $T_c=175$~MeV).}
\vspace{-0.3cm}
\label{fig_spectra}
\end{figure}
Results are shown in Fig.~\ref{fig_spectra}. In the LMR (left panels), 
the strong medium effects in the $\rho$ spectral function 
(cf.~Fig.~\ref{fig_lmr}) entail good agreement with
the CERES data including the low-$q_t$ nature of the excess.  
Within the same fireball model, but using the 'dual' rates from 
Fig.~\ref{fig_imr}, the enhancement observed by NA50 in the IMR 
(1.5~GeV~$\le M\le$~3~GeV) can also be reproduced by thermal 
radiation~\cite{RS99,GKP} (right panels).

\section{Conclusions}
The excess of dileptons measured at low and intermediate masses 
in heavy-ion collisions at the full CERN-SpS energy can be consistently 
attributed to thermal radiation from an expanding fireball.
Whereas the IMR contains a moderate fraction ($\sim$25\%) from  
the highest temperature phases indicative for a QGP, the 
LMR seems to require strong medium effects in the vector channel.
Hadronic model calculations suggest that these might be interpreted
as an in-medium lowering of the quark-hadron 'duality threshold' from
its vacuum value of 1.5~GeV down to $\sim$0.5~GeV. In such a scenario 
chiral restoration is realized by a merging of the
vector and axialvector correlator into their perturbative form.
Advanced investigations of the axialvector channel, as well as the upcoming
measurements at the CERN-SpS and RHIC, are essential for further progress. 

\section*{Acknowledgments}
It is a pleasure to thank my collaborators J. Wambach, E.V. Shuryak, 
G.E. Brown and C. Gale for the fruitful joined efforts. 
This work has been supported in part by the A.-v.-Humboldt foundation
(through a Feodor-Lynen fellowship) and the  U.S. Department of Energy
under Grant No. DE-FG02-88ER40388.

\end{document}